\documentstyle[epsfig]{aipproc}

\begin{document}
\title{Recent Results Addressing the KARMEN Timing Anomaly}

\author{E. D. Zimmerman}
\address{Columbia University, New York, New York}

\maketitle

\begin{abstract}
Recent resuls from experiments at Fermilab and the Paul Scherrer Institute
have constrained the parameter space available for a hypothetical particle
$Q^0$ produced in the decay $\pi^+ \rightarrow \mu^+ Q^0$. This decay has been
invoked to explain a peculiar feature of an event arrival time distribution 
observed in the KARMEN neutrino experiment.
\end{abstract}

\section*{THE KARMEN SIGNAL}

In 1995, the KARMEN collaboration at the Rutherford Appleton Laboratory's
ISIS spallation source published evidence \cite{karmen1} for an 
anomaly in the arrival-time distribution of neutrinos from pion and muon 
decay in a pulsed beam-stop source. The anomaly consisted of an excess of
events with a small amount of electromagnetic energy ($<\sim$35~MeV)
delayed 3.6~$\mu$s with respect to the beam arrival. 

In KARMEN's first data set (1990-95) the anomaly comprised $83\pm 28$
events; data through 1998 \cite{oehler} increased the excess to $103\pm 34$ 
events. 

\subsubsection*{The particle interpretation}

An explanation of the anomaly is that an exotic particle, referred to in
this paper as $Q^0$, is produced by the
rare decay $\pi^+\rightarrow \mu^+Q^0$ near the kinematic threshold for
that process. From a $\pi^+$ decay at rest, the $Q^0$ would travel 
$17.5$~meters to the KARMEN detector in 3.6 $\mu$s (corresponding to 
a velocity $v\approx 0.016 c$). The anomalous signal in the detector 
would be due to the decay of the $Q^0$ to an electromagnetic final 
state. Based on the measured time of flight, the mass of the $Q^0$ must 
be 33.91~MeV. The visible energy of the anomaly favors a three-body
decay to an electron (or possibly photon) pair and an invisible 
final-state particle, likely a neutrino.
A likelihood analysis \cite{oehler} has shown evidence for a correlation 
between position and arrival time for events in the anomaly, consistent with
the slow-moving particle interpretation.

If interpreted as such a particle, the KARMEN signal corresponds to a curve 
in $Q^0$ lifetime versus $BR(\pi^+\rightarrow \mu^+ + Q^0)\cdot 
BR(Q^0 \rightarrow {\rm visible})$. This branching ratio is minimized at
$\sim 10^{-16}$ for a lifetime $\tau_{Q^0}=3.6$~$\mu$s. For any larger 
branching ratio, two solutions exist, one at a longer and one at a shorter
lifetime. A previous experiment
at the Paul Scherrer Institute (PSI) has ruled out at 90\%~C.L any exotic 
$\pi^+$ decays to muons above a branching ratio of 
$2.1\times 10^{-8}$ \cite{daum95}. This constrains the lifetime of the
$Q^0$ to be between $\sim 10^{-7}$ and $\sim 10^3$ s (see Figure~\ref{limit}).

\subsubsection*{Theoretical explanations}

Most theoretical speculation on the anomaly has been along one of
two lines: neutral heavy leptons \cite{govaerts,barger} and light 
neutralinos \cite{choudhury}. Both are discussed briefly below: 

A ``standard'' neutral
heavy lepton (``heavy'' or ``sterile neutrino''), which would be produced
and decayed solely through mixing with $\nu_\mu$, is consistent with the
KARMEN data alone. However, the KARMEN result would require the particle
to have a relatively large mixing with $\nu_\mu$ \cite{glr,shrock}; such a 
large mixing is 
not consistent with the PSI branching ratio limit.  A neutral heavy lepton 
explanation for the $Q^0$ is still allowed if its production is dominated 
by a small mixing with $\nu_\mu$ and its decay is dominated by a much 
larger mixing with $\nu_\tau$. 

Another scenario consistent with the KARMEN data is an $R$-parity violating
neutralino decay. This scenario is allowed in an unconstrained
supersymmetric model, but chargino mass limits \cite{pdg} exclude
a such light neutralino in models such as SUGRA, which introduce 
a chargino-neutralino mass relation.

\section*{FERMILAB E815 (NuTeV)}

Fermilab experiment 815 (NuTeV) has performed a direct search for 
$Q^0$ decay using a beam created by the decays of high-energy pions and 
kaons. The experiment took data during Fermilab's 1996-97 Tevatron fixed
target run, accumulating $2.54\times 10^{18}$ 800~GeV protons on a BeO
target with the detector configured to search for exotic particle decays. 
Sign-selected secondary pions and kaons in the $100-400$~GeV range were 
focused down a 440~m decay pipe, where they could decay before hitting
a steel beam dump. A total of $(1.4 \pm 0.1)\times 10^{15}$ pion decays
occurred in the pipe. Neutral weakly-interacting decay products (neutrinos
and possibly $Q^0$'s) traveled through approximately 900~m of earth berm
shielding before arriving at the detector. 

The E815 detector consisted of an instrumented decay channel followed by
an iron-target neutrino detector. The decay channel was composed 
of an upstream charged particle veto followed by a series of helium bags 
totaling 34~m in length, interspersed with 
3~m~$\times$~3~m multiwire argon-ethane drift chambers. The chambers were 
used to track charged particles from decays in the helium fiducial volume. 
The neutrino detector, which consisted of iron plates interspersed with
drift chambers and liquid scintillator counters, provided calorimetry,
particle identification, and triggering. A toroid spectrometer at the
downstream end of the detector measured the momentum of muons too energetic
to range out in the calorimeter.

In this experiment, the experimental signature of $Q^0$ decay was the 
spontaneous appearence of a low-mass, low-transverse momentum 
electron-positron pair in the helium decay channel. The analysis requirements
were that two well-reconstructed tracks form a vertex in the decay channel
fiducial volume, and that the tracks be identified as electrons based on
the shape of their shower in the calorimeter. Because of the low mass of the
$Q^0$, signal events typically had a very small opening angle and thus 
formed a single merged electron-like cluster in the calorimeter.

Backgrounds to the $Q^0$ search were primarily due to interactions 
between the high-flux neutrino beam and material in the upstream berm,
veto, and decay channel drift chambers. Interactions in the berm and veto
wall could produce photons or neutral kaons which could then convert or
decay to charged particles in the decay channel. Interactions in the 
drift chambers themselves could produce charged tracks directly in the
fiducial volume. These backgrounds were reduced by removing events with
activity in the upstream veto wall, by making tight cuts on electron
identification, and by a series of kinematic cuts designed to discriminate
against the high-$Q^2$ events with large invariant mass final states
typical of neutrino deep inelastic scattering. 

\begin{figure*}
{\center \epsfxsize=6.0in\epsfbox{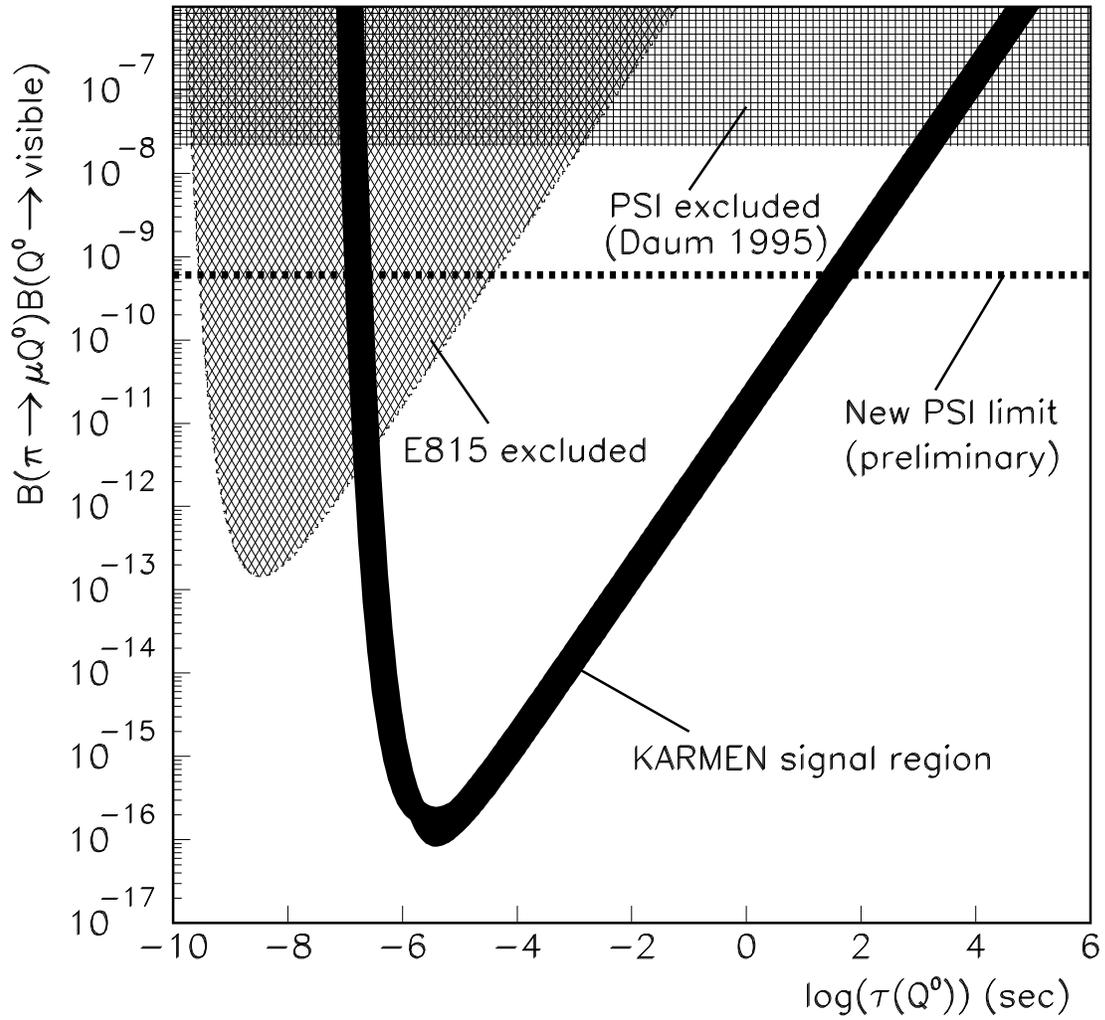} \\}
\caption{The KARMEN signal region and limits from other experiments.\label{limit}}
\end{figure*}

With the final requirements, the expected background level from all sources
was $(0.06 \pm 0.05)$ events with a signal acceptance of 16\% for decays in
the fiducial region. After a blind analysis, the signal region contained 
no events. The result excluded the short-lifetime solution to the KARMEN
anomaly above a branching ratio of $\approx 5\times 10^{-12}$ at 90\% C.L.
(see Figure~\ref{limit} and Ref. \cite{e815prl}).

\section*{THE SEARCH AT PSI}

A new indirect search for $Q^0$ production was recently conducted at PSI by 
M. Daum {\em et al.} \cite{koglin}. The experiment exploited the near-zero
$Q$-value in the $\pi^+\rightarrow \mu^+Q^0$ decay by searching for muons
from $\pi^+$ decay emerging parallel to and with the same velocity as 
the parent $\pi^+$. 

The experiment began with a proton beam aimed at a production target; 
a narrow-band beam of pions with momentum 150 MeV/$c$ was selected and
focused toward a decay region. Downstream of the decay region a beamline
selected charged particles of a particular momentum traveling in the 
forward direction. A series of scintillation counters in the downstream 
beamline measured the velocity of particles in the beamline, allowing 
identification of decay products.

The signature of $Q^0$ production is an excess of forward muons with a 
momentum of 113.5~MeV. The search technique was to scan the analysis
beamline to select different momenta around 113.5~MeV/$c$, in 0.5~MeV/$c$
steps. The number of forward muons was counted in each momentum 
configuration, and the dependence of the muon rate was fit to a $Q^0$ 
plus background distribution. 

The final fit indicated a branching ratio 
$BR(\pi^+ \rightarrow \mu^+Q^0) = (1.3 \pm 2.3)\times 10^{-10}$,
reported as a limit $BR(\pi^+ \rightarrow \mu^+Q^0) < 6.0\times 10^{-10}$
at 95\% C.L.
Because the $Q^0$ decay is not detected, this limit applies to both the
short and long lifetime solutions of the KARMEN anomaly. Unlike the
E815 limit, it is independent of $BR(Q^0 \rightarrow {\rm visible})$.

\section*{STATUS OF THE ANOMALY}

At present, the timing anomaly remains an apparently significant feature
of the KARMEN data set. New data from KARMEN over the next year will be
interesting, but will not be a major addition to the existing data.
Neither the PSI nor the Fermilab groups expect to take more data. 
Their current results are already approaching inherent limitations in their
technique: the unavailability of arbitrarily high beam intensities in the
Fermilab case, and the intrinsic $\pi^+ \rightarrow \mu^+\nu\gamma$ 
background at PSI. Some upcoming experiments, including BooNE (Fermilab 
E898) will have some sensitivity to the $Q^0$, particularly in the
short-lifetime region. A dedicated search for $Q^0$ decay has been proposed
at ISIS, and it is possible that the muon source at a muon storage ring
may be used to search for $Q^0$ production. No experiment currently approved 
is likely to have sufficient sensitivity to confirm or rule out the existence 
of the $Q^0$.

\section*{ACKNOWLEDGMENTS}

The author thanks the E815 (NuTeV) collaboration, M. Daum and J. Koglin 
of PSI, and C. Oehler of the University of Karlsruhe. This work was 
supported by the National Science Foundation.

\end{document}